\documentclass{aastex}

\usepackage{amssymb}
\usepackage{amsmath}

\newcommand{\pa}[2]{\frac{\partial #1}{\partial #2}}
\usepackage{aas_macros}
\usepackage{dsfont}
\newcommand{\dmm}{D_{\mu\mu}}

\newcommand{\cms}{\text{ cm} \cdot \text{s}^{-1}}

\journal{The Astrophysical Journal Supplement Series}

\begin{document}


\title{Determining pitch-angle diffusion coefficients from test particle simulations}

\author{Alex Ivascenko}
\affil{Centre for Space Research, North-West University, 2531 Potchefstroom, South Africa}
\email{alex.ivascenko@rub.de}
\and
\author{Sebastian Lange}
\affil{Lehrstuhl f\"ur Astronomie, Universit\"at W\"urzburg, Emil-Fischer Str. 31, 97074 W\"urzburg, Germany}
\and
\author{Felix Spanier}
\affil{Centre for Space Research, North-West University, 2531 Potchefstroom, South Africa}
\and
\author{Rami Vainio}
\affil{Department of Physics and Astronomy, University of Turku, Finland}

\begin{abstract}
Transport and acceleration of charged particles in turbulent media is a topic of great interest in space physics and interstellar astrophysics. These processes are dominated by the scattering of particles off magnetic irregularities. The scattering process itself is usually described by small-angle scattering with the pitch-angle coefficient $\dmm$ playing a major role.\\
Since the diffusion coefficient $\dmm$ can be determined analytically only for the approximation of quasi-linear theory, the determination of this coefficient from numerical simulations has, therefore, become more important.\\
So far these simulations yield particle tracks for small-scale scattering, which can then be interpreted using the running diffusion coefficients. This method has a limited range of validity.\\
This paper presents two new methods that allow for the calculation of the pitch-angle diffusion coefficient from numerical simulations. These methods no longer analyse particle trajectories, but the change of particle distribution functions. It is shown that they provide better resolved results and allow for the analysis of strong turbulence.\\
The application of these methods to Monte Carlo simulations of particle scattering and hybrid MHD-particle simulations is presented. Both analysis methods are able to recover the diffusion coefficients used as input for the Monte Carlo simulations and provide better results in MHD simulations especially for stronger turbulence.
\end{abstract}

\keywords{diffusion, magnetohydrodynamics (MHD), methods: numerical, scattering, turbulence}



\section{Introduction}
\label{sec:intro}

It is a well-established fact that the transport of energetic charged particles in the heliosphere and the interstellar medium is a convection process along magnetic fields and a diffusive process governed by the collisionless interaction of the particles with magnetic irregularities. A basic concept to describe diffusive particle transport is the Fokker-Planck equation. One of the most important Fokker-Planck coefficients for small-angle scattering is the pitch angle diffusion coefficient $\dmm$. The importance of this parameter lies in its connection to observable quantities and the mean free path of a charged particle in a plasma. The common analytical approaches to derive $\dmm$ use strong approximations, primarily the quasi linear theory (QLT), first suggested by \citet{jokipii66}. The fundamental approximation of the QLT is the assumption of \emph{unperturbed orbits}, meaning that charged particles follow their gyro-trajectories without any disturbance. This is only valid in weakly turbulent plasmas. As the $\delta B/ B_0$ ratio and hence the turbulence strength   increases, particles are scattered significantly and the QLT cannot be applied.

To overcome the problems with analytic approaches numerical simulations have been undertaken. A common approach there is to assume an artificial turbulence spectrum in which tracks of charged particles are followed and the parallel and perpendicular diffusion coefficients are calculated \citep{1997A&A...326..793M,2002ApJ...578L.117Q}. Since spatial diffusion coefficients can be derived from pitch-angle diffusion coefficients and satellite observations yield pitch-angle distribution functions, it is more interesting to use test-particle simulations to derive the pitch-angle diffusion coefficient directly \citep{2012ApJ...750..150W}.

For a number of reasons, the derivation of the pitch-angle diffusion coefficient is far more complicated than the derivation of spatial diffusion coefficients. One of the main reasons is the changing pitch-angle when tracing particle tracks in real turbulence, which leads to a failure of the classical method of running diffusion coefficients since the pitch-angle diffusion coefficient has to be calculated per pitch-angle. This has also been shown by \cite{qinshalchi2009}, who conclude that due to the rapidly changing $\mu$ in strong turbulence the pitch-angle diffusion coefficient cannot be determined with the running diffusion method. Despite the limitations of the method, it has nevertheless been successfully used to derive pitch-angle diffusion coefficients from numerical simulations and gain insights into the nature of diffusive particle transport, e.g. scattering at $\mu=0$ \citep{qinshalchi2014} and subdiffusion in 2D-turbulence \citep{qinshalchi2009}.

In this paper we present new approaches to calculate $\dmm$ and compare them to the QLT derivations. These methods are distinguished from common derivations of $\dmm$ by their applicability independent of the $\delta B/ B_0$ ratio.

\section{Theory}
\label{sec:theory}

\subsection{Particle transport basics}
\label{sec:statistictransport}

Particle transport in turbulent media is a stochastic process, which is typically described using a statistical approach. An extensive discussion of the foundations can be found in \citet{Schlickeiser2002}. For our study the derivation of the transport theory from the Vlasov equation is not especially important, we will focus on the Fokker-Planck equation, which can be derived from the \emph{quasilinear theory}, first suggested by \citet{jokipii66} in the context of energetic charged particle transport in turbulent magnetic fields. The fundamental assumption is that of unperturbed particle orbits. This implies the fluctuation amplitudes to be small, leading to a quasilinear system. The Vlasov equation for the particle distribution function $F_T$ then simplifies to \citep{Schlickeiser2002}
\begin{align}
 \pa{F_T}{t} + v\mu \pa{F_T}{Z} &- \Omega \pa{F_T}{\phi} = S_T(X_\sigma,t) \nonumber  \\
 &+ \frac{1}{p^2} \pa{}{X_\sigma} \left(  p^2 \pa{F_T}{{X_\eta}} \underset{D_{X_\sigma {X_\eta}}}{\underbrace{\int_0^t \text{d}s \langle g_{X_\sigma}g_{{X_\eta}}({X_\eta},s) \rangle}} \right).
\label{eq:final_FokkerPlanck}
\end{align}
This equation is known as the Fokker--Planck equation with the Fokker--Planck coefficients $D_{X_\sigma {X_\eta}}$, where the $X_\sigma, X_\eta$ are generalized coordinates. One of the most interesting parameters is the pitch--angle diffusion coefficient $D_{\mu\mu}$. It describes the pitch angle scattering of the particle and is consequently connected to the scattering mean free path, which can be evaluated by the observable angular distribution and particle transport simulations \citep{2009AguedaVainio}.

When $D_{\mu\mu}$ is the dominant component, one may simplify the Fokker-Planck equation to
\begin{align}
  \pa{F_T}{t} + v\mu \pa{F_T}{Z} &- \Omega \pa{F_T}{\phi} = \pa{}{\mu}\left(D_{\mu\mu} \pa{}{\mu} F_T\right).
\end{align}
This in turn can be further simplified after averaging over phase angles $\phi$ and space to
\begin{align}
    \pa{F_T}{t} = \pa{}{\mu}\left(D_{\mu\mu} \pa{}{\mu} F_T\right) \label{eq:diffapprox}.
\end{align}

In this context, scattering represents a resonant wave--particle interaction of the $n$th order which fulfills the condition
\begin{align}
 k_{\parallel} \,  v_{\parallel} - \omega + n \, \Omega = 0, \quad n \in \mathds{Z}
\label{eq:wave-particle-res}
\end{align}
(cf. \citet{schlickeiser89}), where $\omega$ is the wave frequency and $ k_{\parallel}$ its parallel wavenumber. $\Omega$ is the gyro-frequency and $v_{\parallel}$ its parallel velocity component. Different components of the waves contribute for certain values of $n$. Namely, the Cherenkov resonance with $n=0$ is generated by either compressible waves or by pseudo Alfv\'en waves in the incompressible regime through the mirror force induced by magnetic compressions. Shear Alfv\'en waves with a wave vector strictly parallel to the background magnetic field ($k_\perp = 0$) fulfill the resonance condition Eq. \ref{eq:wave-particle-res} only for $n=\pm 1$, while for $k_\perp \neq 0$ resonances with $|n|>1$ are possible (as a result of non-vanishing higher order Bessel functions in the derivation).

\subsection{Particle Transport: Numerical approach}
\label{sec:numerics}

We use two different numerical approaches in this paper to simulate particle scattering: The first method is based on a Monte Carlo approach, which produces pitch-angle distributions using a prescribed diffusion coefficient, the second method calculates particle scattering from the interaction of charged particles with a turbulent background.\\
The reason to use this two-fold approach is that the first method may used to validate the results, since we may compare the results with a given input parameter, while the second method is a typical use case. This use case shows a far more realistic spectrum for  turbulence.

\subsubsection{Validation using Monte Carlo methods}

To validate the results of the new diffusion coefficient calculation method it was applied to the output of a Monte Carlo  propagation code with a given pitch angle diffusion coefficient as input. For a detailed description of the code see \cite{agueda2008}, \cite{agueda2013}.

For the validation of method MII it was necessary to use a particle distribution, which is not zero anywhere ($f(\mu)\neq 0$), where the derivative is not vanishing ($df/d\mu \neq 0$), and which is invertible in $\mu$ initially. The choice is otherwise arbitrary. The Monte Carlo method is not affected by any choice of the distribution function.

\subsubsection{Application to MHD turbulence}

In order to investigate particle transport in a turbulent plasma, a numerical approach has been chosen. For this purpose, the parallel hybrid--code \textsc{Gismo} was developed, which consists of two parts. GISMO solves the incompressible MHD equation using a pseudospectral method and traces the motion of charged test particles, which interact with the electromagnetic fields generated by the MHD turbulence in the plasma. A short description is given in Appendix \ref{app:gismo}. For a detailed description we refer to \citet{lange2012}.\\
The basic setup uses an anisotropic turbulence, which is driven continuously by injecting energy into certain wave modes.
The magnetic background field in the first simulation setup is approximately $B_0 = 0.174 \text{ G}$, which yields, assuming a particle density of $10^5 \text{ cm}^{-3}$, an Alfv\'en speed of $v_A = 1.2 \cdot 10^8 \text{ cm}\text{ s}^{-1}$ .
These values resemble conditions in the solar corona at a distance of three solar radii \citep{ramigramm}. The outer length scale of the simulated system is $L_\text{scale}=3.4\cdot 10^8 \text{cm}$. Wave numbers are given in terms of the normalized wave number $k' = k L$. Simulations have been performed on a $256^3$ grid.\\
We have used different turbulence setups as described in \citet{lange2012} and \citet{2013A&A...553A.129L}:
\begin{itemize}
 \item[(1)] A turbulence simulation with anisotropic driver at small wave numbers up to a saturated turbulent stage
 \item[(2)] The same turbulence with an amplification of the wave mode at $k'_\parallel = 2\pi \cdot 24$ (further called peak simulation) during driving stage at small amplitudes
 \item[(3)] The peak simulation at the decay stage with QLT compatible amplitudes
 \item[(4)] The peak simulation at maximum driven stage with big amplitudes, where assumptions of QLT are not fulfilled anymore
\end{itemize}

Into the turbulent plasma test particles are injected - at least $10^6$ particles.
The proton speed was set to a value of $1.21 \cdot 10^{10} \cms$, which was chosen to fulfill the resonance condition (Eq. \ref{eq:wave-particle-res}). Consequently, a resonant value of $\mu$
\begin{align}
 \mu_R = \frac{\omega-n\,\Omega}{k_\parallel\, v} = \frac{\omega-n\,\Omega}{L_\text{scale}^{-1} \,k'_\parallel\, v}
 \label{eq:mures}
\end{align}
must be within the interval $[-1,1]$ for the given particle speed $v$ and wave frequency $\omega$.\\
The pitch-angle distribution does not affect the plasma dynamics as there is no back-reaction of the particles to the plasma induced.  The initial distribution in $\mu$ of the test particles is not important for method MI, only a sufficiently high particle number is needed since this method is statistical. However, the methods of MII depend on the derivative in $\mu$, which means a significant change to the initial distribution. A half--parabola ($f(\mu)=a\cdot(\mu\pm 1)^2 +c$) distribution was, therefore, chosen to achieve a non-zero and non-constant derivative.

\section{Methods to derive the pitch-angle diffusion coefficient}
\label{sec:methods}

In section \ref{sec:statistictransport} the fundamental description of the diffusion coefficient $\dmm$ was given. This section focuses on different concepts to derive $\dmm$ from numerical simulations. Starting with common QLT approaches, we present new methods afterwards.\\
MI is an established method, that relies on the analysis of single particle tracks, and serves as a benchmark for the new methods MIIa and MIIb. Both new methods have in common that they use the particle distribution function for the calculation.

\subsection{MI Running diffusion coefficient}

A simple approach for calculating the pitch angle scattering coefficient is the definition
\begin{align}
D_{\mu\mu} =  \lim_{t \to \infty} \frac{(\Delta \mu)^2}{2 \, \Delta t} \stackrel{t\gg t_0}{\approx} \frac{(\Delta \mu)^2}{2 \, \Delta t},
\end{align}
where $\Delta t = t - t_0$ is assumed to be large, i.e. the time evolution $t$ has to be sufficient to develop resonant interactions.
This approach is motivated by a description of diffusion, where a particle changes its pitch angle by scattering in a randomized process.
If the scattering is in resonance with a wave mode, $\Delta \mu$ increases significantly. This method predicts a $\delta$--function shape in the limit of infinite time development.
However, in finite intervals of $\Delta t$ the resonances are always broadened. Another problem is the dependence on the strength of the scattering process.
In the case of high $\delta B / B_0$ ratios and thus high scattering frequencies, the pitch angle at time $t$ is not connected to its initial state anymore and the scattering coefficient becomes unstructured \citep{2013A&A...553A.129L}.

\subsection{MIIa Diffusion equation fitting method}

A completely different approach is the calculation via the diffusion equation. The basic concept is the assumption of a diffusion process, where the pitch angle diffusion is the predominant process (cf. Eq. \ref{eq:diffapprox})
\begin{align}
 \pa{f_T}{t} - \pa{}{\mu} \dmm \pa{f_T}{\mu} = 0.
 \label{diffeq}
\end{align}
This allows us to calculate the diffusion coefficient from the static particle distribution in $\mu$-space at two distinct timestamps by solving the diffusion equation
\begin{equation}
    \frac{\partial f_T(\mu,t)}{\partial t} = \left(\frac{\mathrm{d}}{\mathrm{d} \mu}D_{\mu \mu}(\mu)\right) \cdot \frac{\partial f_T(\mu,t)}{\partial \mu} + D_{\mu \mu}(\mu) \cdot \frac{\partial^2 f_T(\mu,t)}{\partial \mu^2}
    \label{diffgl}
\end{equation}
numerically for $D_{\mu \mu}(\mu)$.\\
Since the simulations provide us with discrete distributions, the derivatives are discretized accordingly in the usual way, yielding an equation for every $\mu^n = -1 + n\cdot \Delta \mu$ (with $D_{\mu \mu}^n=D_{\mu \mu}(\mu^n)$):
\begin{equation}
    \partial_t f = \frac{D_{\mu \mu}^{n+1} - D_{\mu \mu}^{n-1}}{2\cdot \Delta \mu} \partial_{\mu} f +D_{\mu \mu}^n \partial_{\mu \mu} f
    \label{diskret}
\end{equation}
This corresponds to a matrix equation with a tridiagonal matrix which can be solved with conventional algorithms:
\begin{equation}
    \begin{pmatrix}
        \partial_{\mu \mu} f^0 & \frac{\partial_{\mu} f^0}{2 \Delta \mu} & 0 & 0 \\
        -\frac{\partial_{\mu} f^1}{2 \Delta \mu} & \partial_{\mu \mu} f^1 & \ddots & 0      \\
        0      & \ddots & \ddots & \frac{\partial_{\mu} f^{n-1}}{2 \Delta \mu} \\
        0      &   0    & -\frac{\partial_{\mu} f^n}{2 \Delta \mu} & \partial_{\mu \mu} f^n
    \end{pmatrix}
    \cdot
    \begin{pmatrix}
        D_{\mu \mu}^0 \\ D_{\mu \mu}^1 \\ \vdots \\ D_{\mu \mu}^n
    \end{pmatrix}
    =
    \begin{pmatrix}
        \partial_t f^0 \\ \partial_t f^1 \\ \vdots \\ \partial_t f^n
    \end{pmatrix}
    \label{matrix}
\end{equation}
A problem of this method is the imperfect sampling of the phase space with the test particle approach, which results in a rather noisy distribution function and even noisier derivatives. This can be handled by averaging over several simulation runs, applying smoothing algorithms or fitting the data with analytical functions.\\
While the ensemble averaging is the correct way to increase the signal-to-noise ratio, it would mean a huge computational effort to obtain more simulation data and can't be applied at all to real measurements. We found that fitting sub-sets of the distribution function with low-degree polynomials as described by \citet{savgol64} is the best way to reproduce the main features of the distribution while reducing noise to a manageable level. An additional advantage of the Savitzky-Golay method is the ease to obtain derivatives.

\subsection{MIIb Diffusion equation integration method}

An additional way to deal with the noisy derivatives is to integrate the diffusion equation numerically over $\mu$
\begin{equation}
    \int_{-1}^{\mu} \frac{\partial f_T(\mu,t)}{\partial t} \mathrm{d} \mu = D_{\mu \mu}(\mu) \frac{\partial f_T(\mu,t)}{\partial \mu} = -j_{\mu} (\mu)
    \label{integr}
\end{equation}
thus gaining the effective pitch angle current $j_{\mu}$ that yields the diffusion coefficient when divided by $\partial_{\mu} f_T$. The advantage of this method is that the time derivative of $f_T$ is smoothed by the integration and we only need the first derivative in $\mu$, which can also be approximated by a polynomial if necessary.

\section{Results}

\subsubsection*{Monte Carlo verification}

\begin{figure}[ht]
  \begin{center}
      \includegraphics[width=1. \columnwidth]{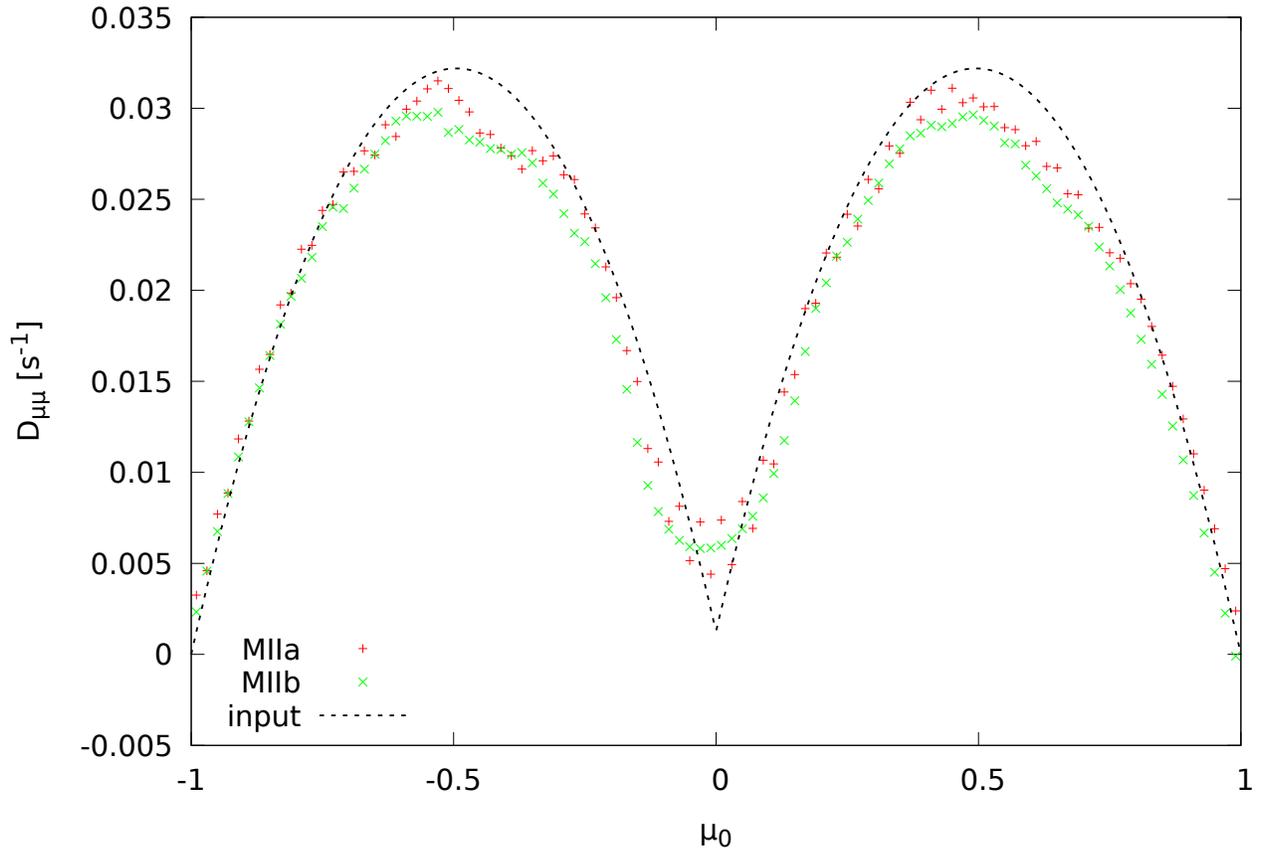}
      \caption{Comparison of $\dmm$ used as input for the Monte Carlo propagation model and results obtained from the output with the newly developed methods MIIa and MIIb.}
      \label{fig:ramiDmm}
  \end{center}
\end{figure}

We first present a validation of the new methods MIIa and MIIb by applying them to the output of a Monte Carlo propagation model with a preset pitch angle diffusion coefficient of the form
\[ \dmm(\mu) = \frac{\nu_0}{2}\left(\frac{|\mu|}{1+|\mu|}+\epsilon \right) \left( 1-\mu^2 \right). \]
The results in Fig. \ref{fig:ramiDmm} show a good agreement in shape and absolute value of the diffusion coefficient calculated from (smoothed) pitch angle particle distributions as compared to the analytical expression used as input.

\subsubsection*{Background simulation results}

For the analysis of MHD simulation results with the new methods we first apply them to so-called background simulations. These are simulations performed with the GISMO code, where energy is injected into an MHD plasma continuously until a steady turbulent spectrum evolves. The Goldreich-Sridhar like spectrum can be seen in Fig. \ref{fig:gismoback}. The fluctuation level is rather low ($(\delta B/B_0)^2\approx 0.001$), which means that the quasi-linear theory is a relatively good approximation for the transport of charged particles in this case.

\begin{figure}[ht]
  \begin{center}
          \includegraphics[width=1. \columnwidth]{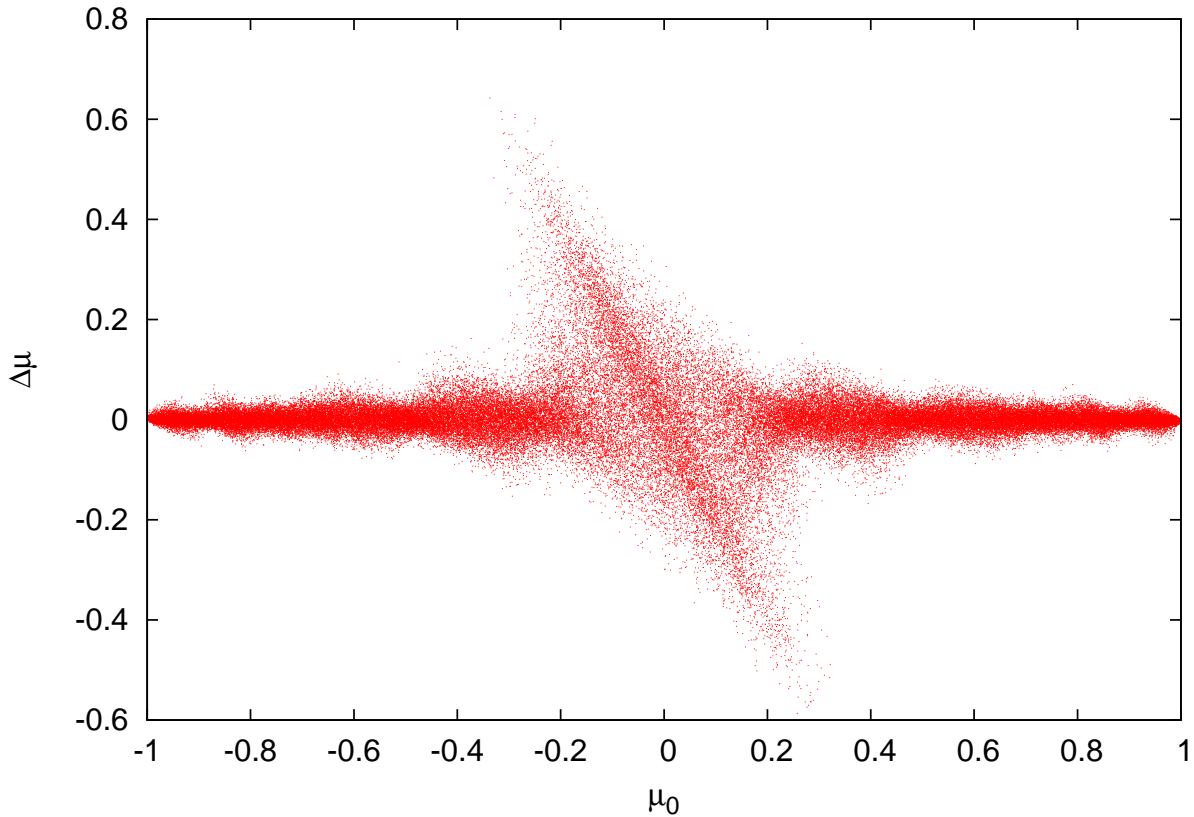}
          \caption{Change of the pitch angle $\Delta \mu$ in dependence of the initial value $\mu_0$ for $10^5$ particles within a turbulent plasma after 30 gyrations. This \emph{scatter plot} is a valuable tool to investigate resonant interactions. The dominant structure is the Cherenkov resonance $n=0$.}
          \label{fig:v31-background-deltamue-30gyr}
  \end{center}
\end{figure}

In \citet{lange2012} scatter plots have been used extensively to display the effect of the diffusion. One such scatter plot is shown in Fig. \ref{fig:v31-background-deltamue-30gyr}. It directly shows the change of the pitch angle $\Delta \mu$ for each individual particle. This type of plot reveals resonant structures very clearly. Unfortunately, this plot is an appropriate tool only for approaches where the individual particle can be traced.

\begin{figure}[ht]
  \begin{center}
          \includegraphics[width=1. \columnwidth]{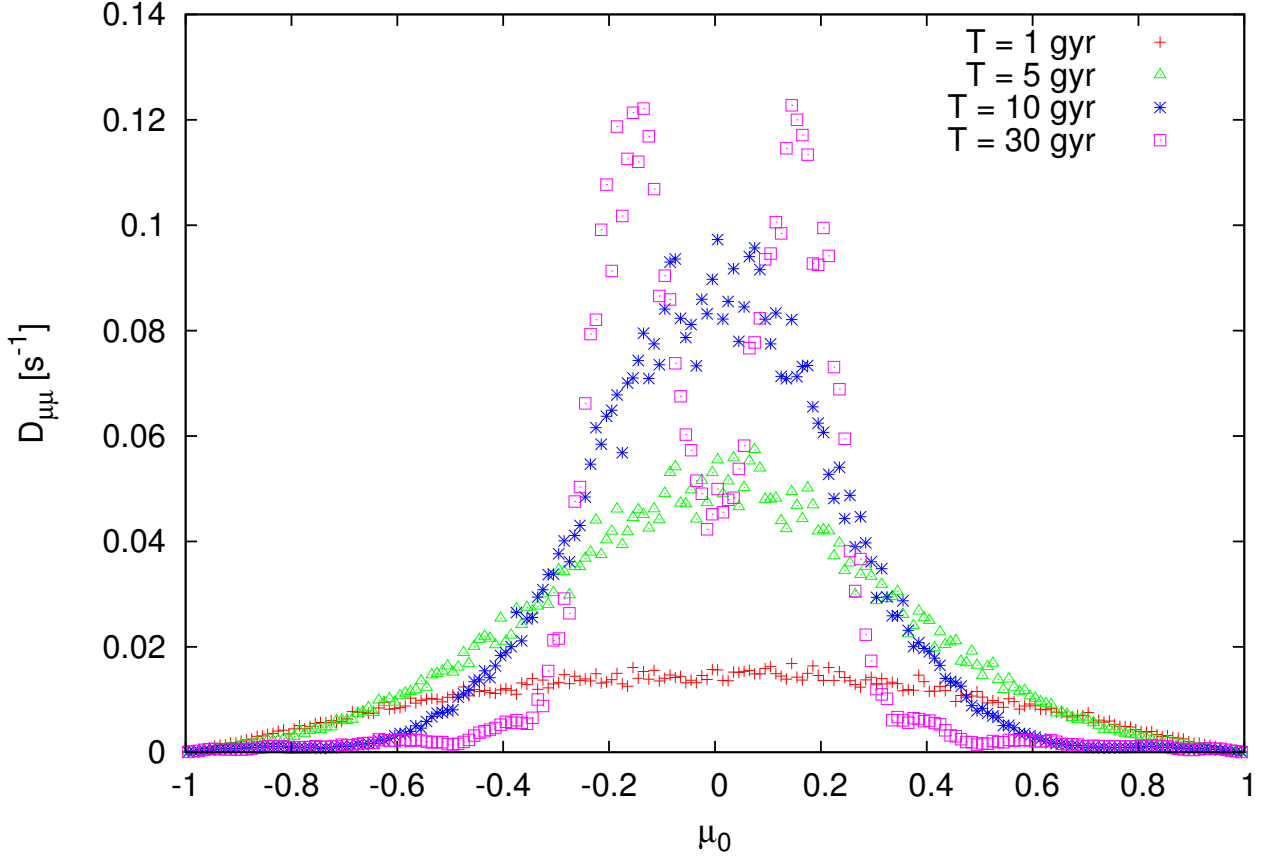}
          \caption{Time evolution of the pitch angle scattering coefficient $\dmm$ calculated by MI. A clear resonant structure develops between 10 and 30 gyration periods. By the comparison to the scatter plot Fig. \ref{fig:v31-background-deltamue-30gyr} the maxima can be connected to the Cherenkov resonance.}
          \label{fig:v31-backgroundwithoutpeak-Dmm}
  \end{center}
\end{figure}

Fig. \ref{fig:v31-backgroundwithoutpeak-Dmm} shows the diffusion coefficient $\dmm$ obtained with the classical running coefficient method MI. The development of a resonant structure is clearly visible as the simulation progresses from 1 to 30 gyration periods. It should be noted that this development is a stochastic effect of an increasing portion of homogeneously distributed particles undergoing resonant interactions with wave modes running through the simulation box and is unrelated to the development of the turbulence itself which is completed before test particles are injected into the simulation.\\
The apparent splitting of the maxima at 30 gyrations is caused by the tilt of the resonance peak in $\Delta \mu$ when plotted over the initial $\mu_0$ as seen in Fig. \ref{fig:v31-background-deltamue-30gyr}. Calculating $(\Delta \mu)^2$ folds the negative half-peak up resulting in the apparent double-peak structure. The correct choice of the starting pitch angle $\mu_0$ has been discussed by \citet{2013A&A...558A.147T}. While we agree that this can improve the appearance of the plot, it does not change its fundamental meaning.

\begin{figure}[ht]
  \begin{center}
          \includegraphics[width=1. \columnwidth]{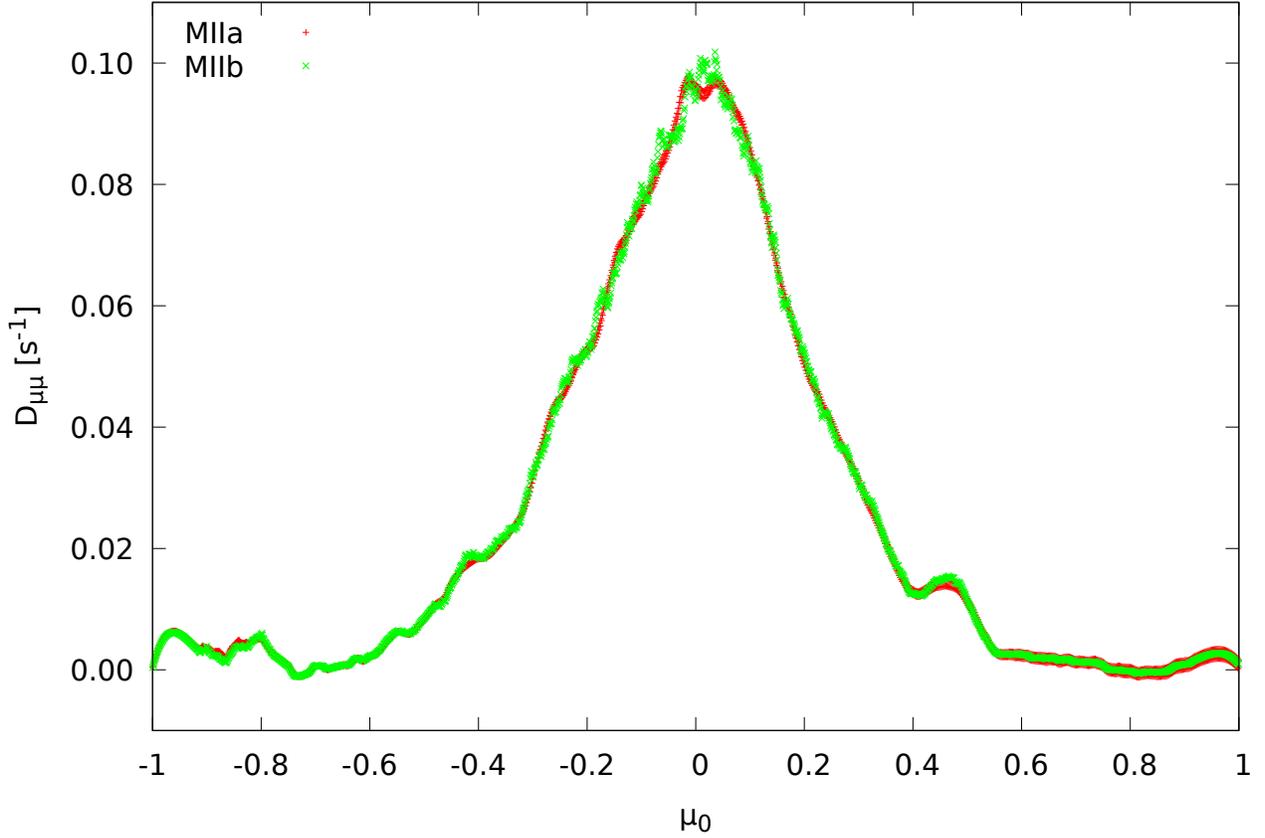}
          \caption{Comparison between MIIa and MIIb for the background turbulence simulation clearly showing the Cherenkov resonance. The shown time interval is after 10 gyration.}
          \label{fig:v31-backgroundwithoutpeak-rabenDmm}
  \end{center}
\end{figure}

The results of the direct integration MIIb in Fig. \ref{fig:v31-backgroundwithoutpeak-rabenDmm} compare very nicely with the direct solution of the fitted matrix equation MIIa. Both methods of MII show a maximum at the Cherenkov resonance which is dominating the scattering (see Fig. \ref{fig:v31-background-deltamue-30gyr}). Unlike MI the maximum does not split due to the tilt shown in the scatter plot. Thus, MII can be used independently and without any interpretative help.

\subsubsection*{Driven peak simulation results}

\begin{figure}[ht]
  \begin{center}
          \includegraphics[width=1. \columnwidth]{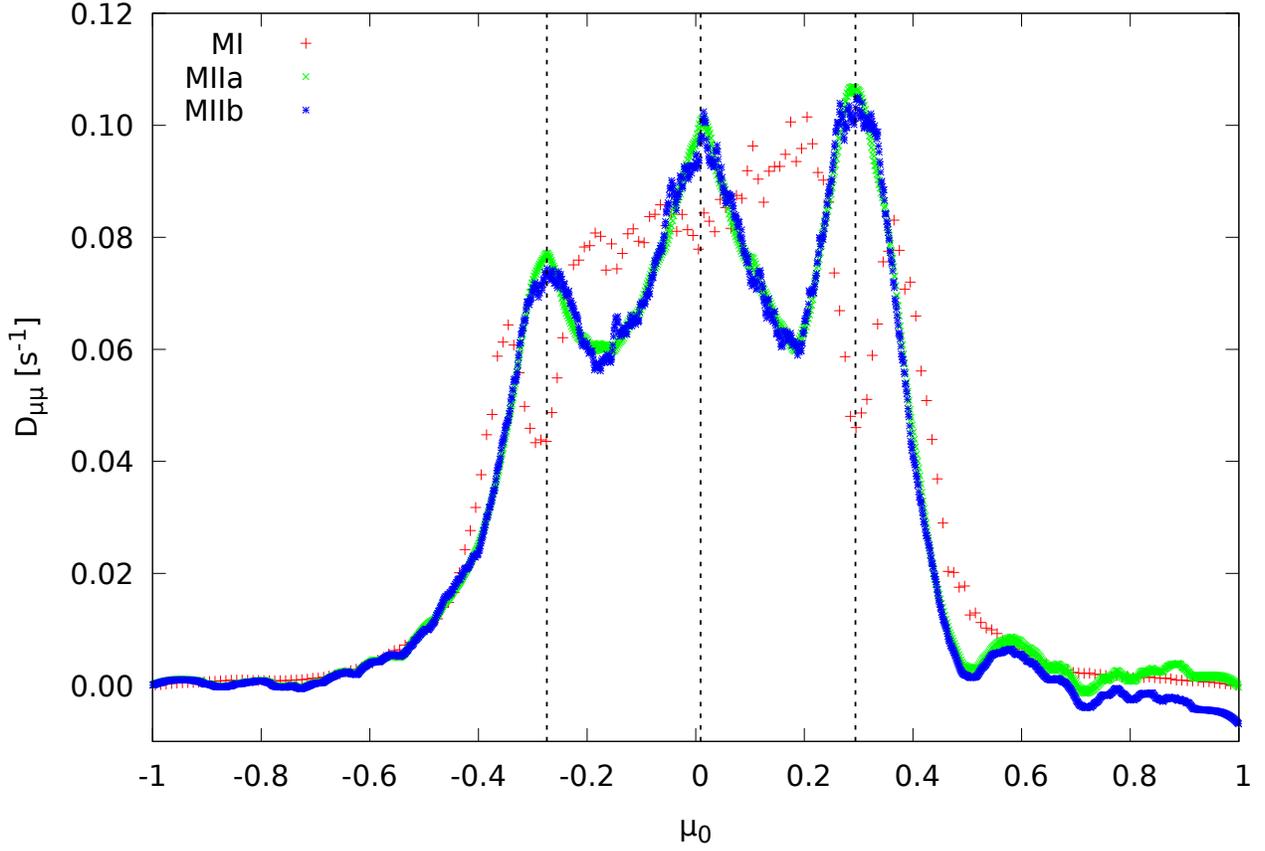}
          \caption{Comparison between MI, MIIa and MIIb for the driven stage of the peak simulation at small amplitudes. The vertical lines indicate the resonance positions according to Eq. \ref{eq:wave-particle-res}. The shown time interval is after 10 gyrations.}
          \label{fig:v31-peak24-drv-halbparabel-rabenDmm}
  \end{center}
\end{figure}

The next scenario is the peak simulation with a driven amplification at $k'_\parallel = 2\pi \cdot 24$. In this scenario a background MHD simulation is used, where additional energy is injected at $k'$. This leads to a strong magnetic fluctuation at localized wave modes. This in turn means that the QLT approximation for the particle transport does not hold anymore at those wave modes. This is an interesting test case for the particle transport analysis since QLT may or may not be applicable depending on the time since the onset of peak driving and the particle energy.

First, we present the results at the beginning of the driven stage, where amplitudes are small enough to keep near the quasilinear assumptions. A comparison between MI and MIIa as well as MIIb is presented in Fig. \ref{fig:v31-peak24-drv-halbparabel-rabenDmm}.
The running diffusion coefficient method MI shows again a split of the maxima, so that the predicted resonant positions are right in between.
Although this is not correct, it is still interpretable by using the corresponding scatter plot. The methods of MII, however, show again a very nice match to the predicted resonances. The small fluctuations at $\mu=0.6, 0.8$ and 0.9 are not connected to resonances, but indicate the statistical influence of the particle number, which in this simulation is lower at positive pitch angles. All of the three methods show comparable amplitudes in $\dmm$.

\subsubsection*{Decaying peak simulation results}

\begin{figure}[ht]
  \begin{center}
          \includegraphics[width=1. \columnwidth]{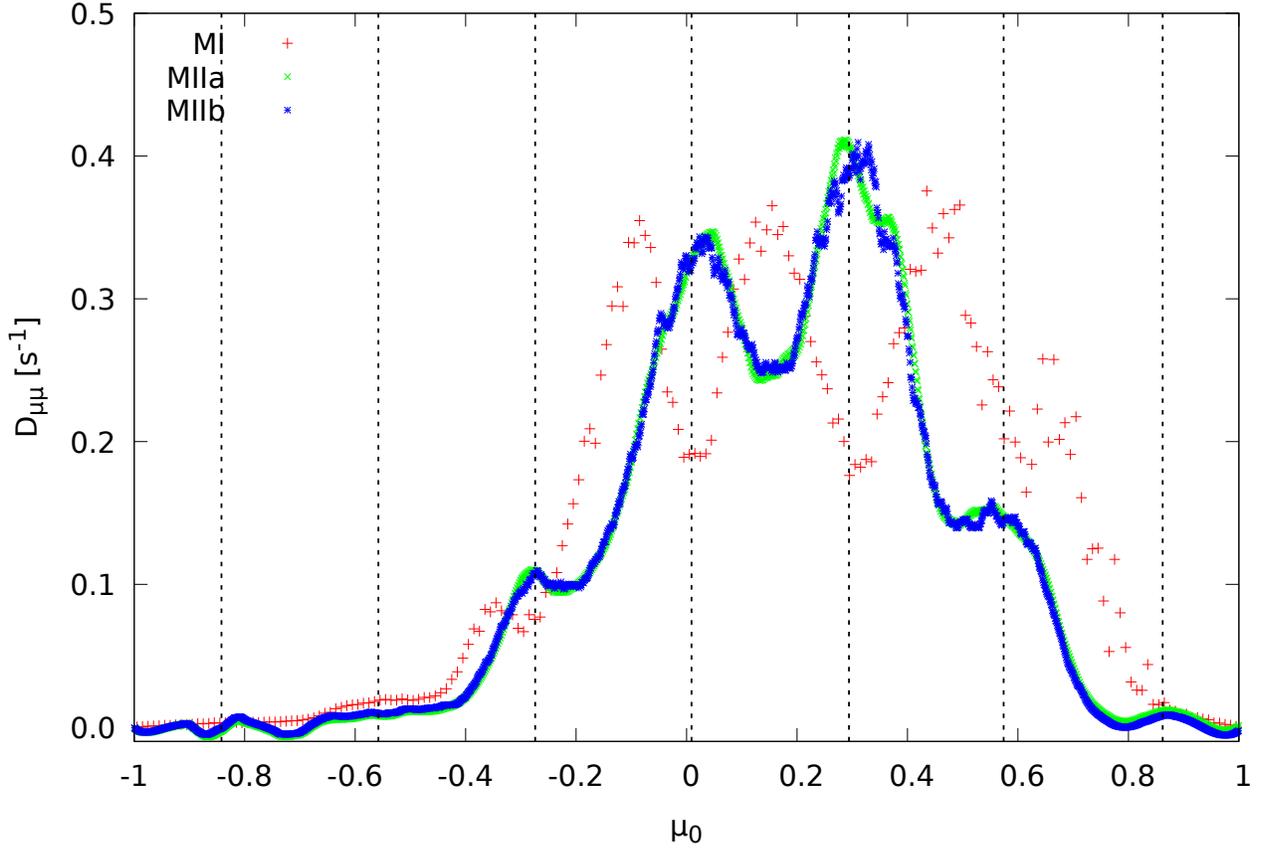}
          \caption{Comparison between MI, MIIa and MIIb for the decay stage of the peak simulation at small amplitudes. The vertical lines indicate the resonance positions according to Eq. \ref{eq:wave-particle-res}. The shown time interval is after 10 gyrations.}
          \label{fig:v31-peak24-dec-halbparabel-rabenDmm}
  \end{center}
\end{figure}

During the decay stage of the peaked mode energy spreads due to convection and diffusion towards perpendicular wave numbers. This leads to increased influence of higher order resonances and thus a more complex scattering pattern. In Fig. \ref{fig:v31-peak24-dec-halbparabel-rabenDmm} we present again the comparison of the results of the different methods during the decay stage of the peak with small amplitudes.

\begin{figure}[ht]
  \begin{center}
          \includegraphics[width=1. \columnwidth]{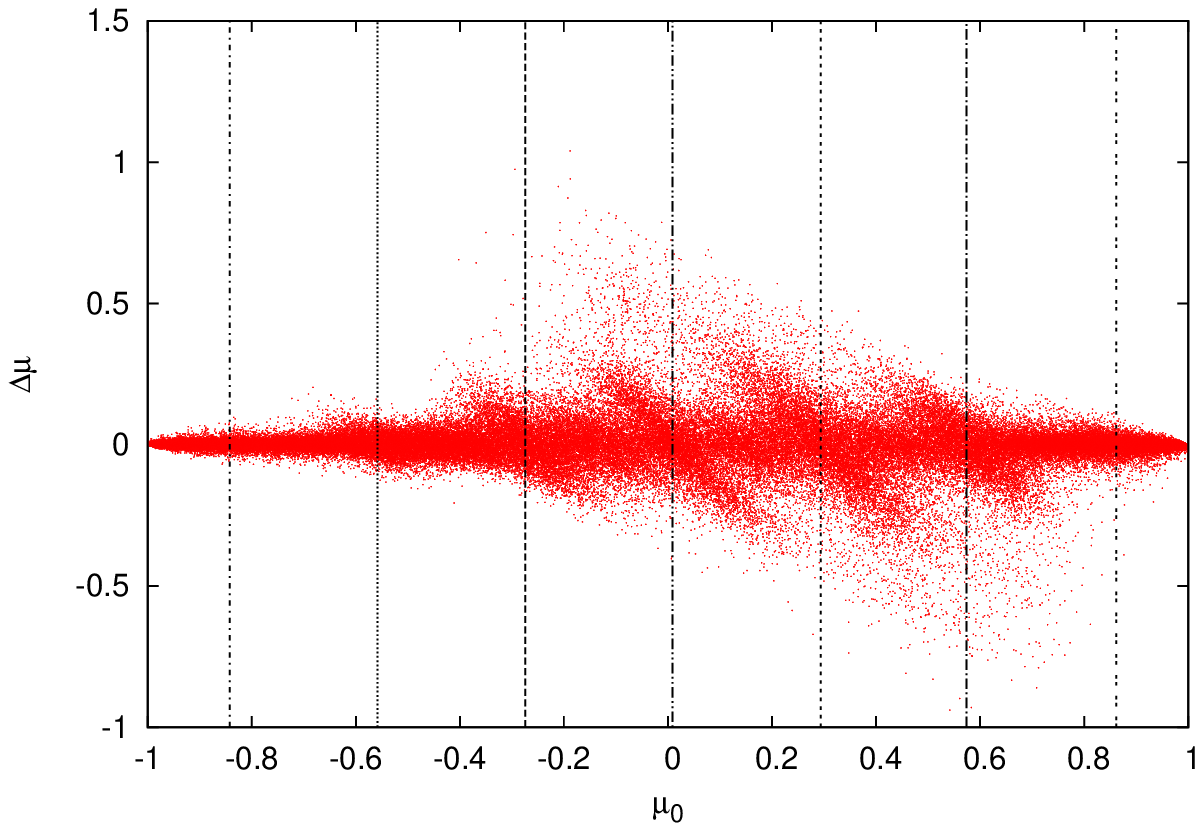}
          \caption{Scatter plot of the peak resonance during the decay stage.}
          \label{fig:v31-small24-decay-deltamue}
  \end{center}
\end{figure}

Because the scattering is more complex, we present also the corresponding scatter plot in Fig. \ref{fig:v31-small24-decay-deltamue} to interpret the results of MI. The tilt of the resonances causes again the split of the maxima in the results of MI. The left hand polarization of the wave mode has not developed and resonances with negative $\mu$ are smaller compared to $n=2$ and 3. MIIa and MIIb confirm this.

\subsubsection*{Strong turbulence results}

\begin{figure}[ht]
  \begin{center}
          \includegraphics[width=1. \columnwidth]{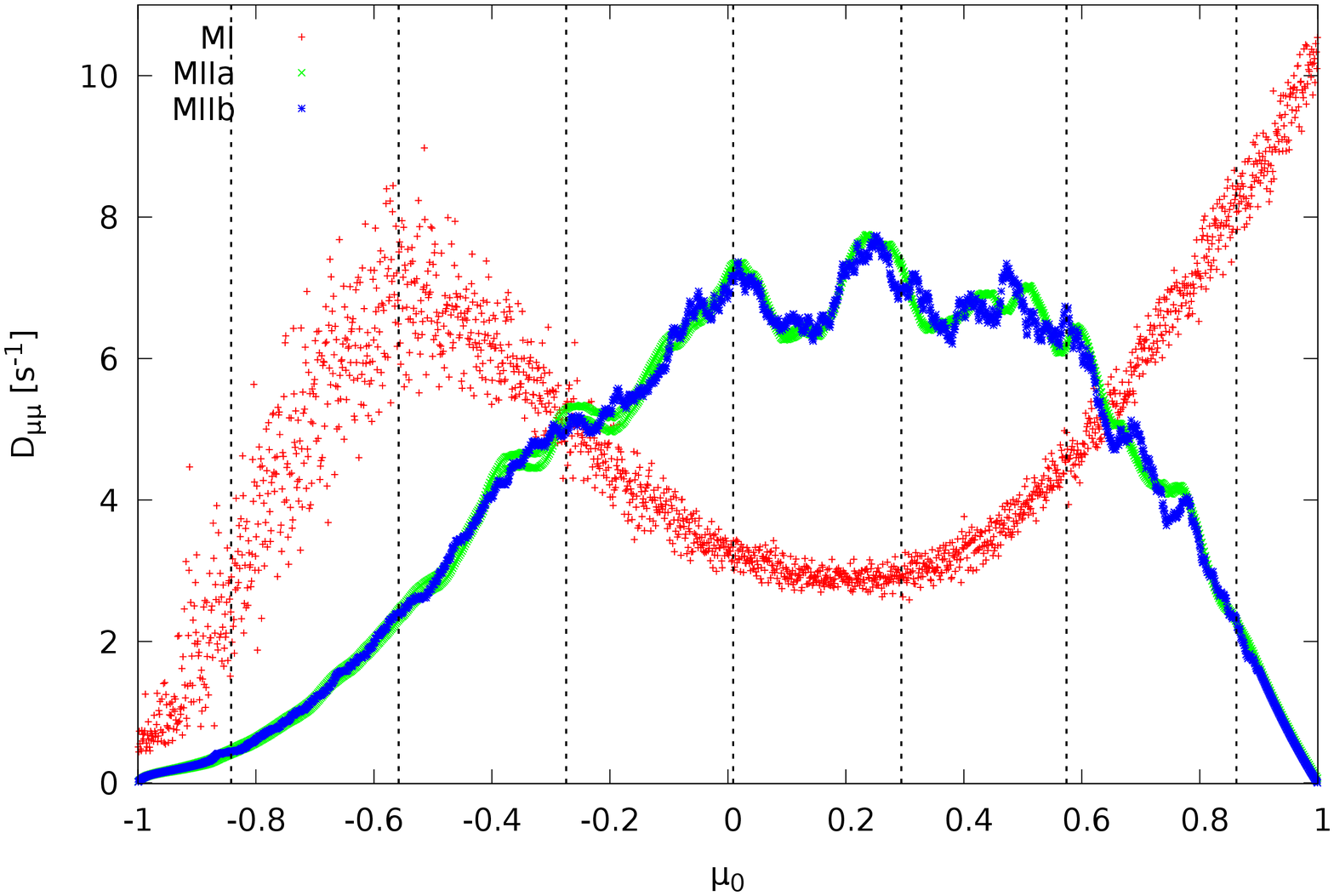}
          \caption{Comparison between MI, MIIa and MIIb for the maximum driven stage of the peak simulation at big amplitudes. The vertical lines indicate the resonance positions according to Eq. \ref{eq:wave-particle-res}. The shown time interval is after 10 gyrations. The QLT based MI derives an incorrect $\dmm$, the match with the resonance at $\mu=-0.56$ is coincidental. Both methods of MII show the maxima at the predicted positions of the resonances.}
          \label{fig:v31-peak24-maxdrive-halbparabel-rabenDmm}
  \end{center}
\end{figure}

In order to test the limits of QLT we present results of the peak simulation at maximum driven stage. The local amplitudes reaches values of $\delta B / B_0 \approx 1$ where quasilinear assumptions are not valid. Consequently, the results in Fig. \ref{fig:v31-peak24-maxdrive-halbparabel-rabenDmm} of method MI are erroneous and also not interpretable with the scatter plot anymore. The corresponding scatter plot (not shown here) only shows a broad tilted band where many particles are scattered strongly. Nevertheless, the methods of MII are expectedly not influenced by this. Both results show again the resonant maxima, although there is a larger background scatting.

\section{Conclusion}

We have presented two different methods to calculate the Fokker-Planck coefficient $\dmm$. The first method MI depends on the QLT and is thus sensitive to turbulence strength or wave mode amplitudes. Consequently, this method is usable for very small fluctuations $\delta B$ and $\delta E$ only.
Despite this sensitivity to the validity of the QLT assumption, as shown in our results, the method MI is still applicable if corresponding scatter plots are used for interpretation. Only at big wave amplitudes this method fails (see Fig. \ref{fig:v31-peak24-maxdrive-halbparabel-rabenDmm}).

The methods of MII are both independent of quasilinear assumptions. This is very important for most scenarios.
Especially with particle scattering in highly turbulent states or at wave modes with large amplitudes these methods give correct results.
The resonant maxima are not split up and interpretation with the scatter plots is not necessary, which is another advantage.
However, care should be taken considering the assumptions incorporated in the new methods. Since Eq. \ref{diffeq} presupposes a purely diffusive process in $\mu$, MII can't be used directly in case of anomalous diffusion, contrary to MI, which can be employed to calculate diffusion coefficients in super- and subdiffusion \citep{qinshalchi2009}. The integer derivatives in Eq. \ref{diffeq} change to fractional derivatives in this case, resulting in the equation
\begin{align}
    \pa{f_T}{t} - D_{\alpha} \mathbb{D}^{1-\alpha}_t \frac{\partial^2 f_T}{\partial \mu^2} = 0
\end{align}
with the fractional derivative definition
\begin{align}
    \mathbb{D}_t^{1-\alpha} = \frac{1}{\Gamma(\alpha)} \pa{}{t} \int_0^t \frac{f_T(t')}{(t-t')^{1-\alpha}} dt'.
\end{align}
While an inversion of this fractional derivative exists mathematically, the actual implementation would involve solving Laplace integrals, making the method even more sensitive to noise and destroying the simple and elegant form of MII.

A different approach could be taken to tackle anomalous diffusion in an approximate manner by calculating $\dmm(t) \propto t^{\alpha-1}$ at successive timesteps with unmodified Eq. \ref{diffeq} and determining the fractional order $\alpha$ from it. Unfortunately, to have a chance to see anomalous diffusion in our MHD simulations would require large simulation sizes (for superdiffusion) and long run times (for subdiffusion).

Another assumption of MII concerns the pitch angle distribution function $f_T(\mu)$. It must have a non-zero first and second derivative in $\mu$ and should change sufficiently between two evaluation timesteps.
Consequently resonances with very small wave amplitudes or those not leading to a change in the $\mu$-distribution are not resolved. In this case MI should be used.

\section{Acknowledgements}

AI and FS acknowledges support from NRF through the MWL program. This work is based upon research supported by the National Research Foundation and Department of Science and Technology. Any opinion, findings and conclusions or recommendations expressed in this material are those of the authors and therefore the NRF and DST do not accept any liability in regard thereto.

\appendix

\section{Description of the GISMO code}
\label{app:gismo}
The GISMO code \citep{lange2012} was used to determine turbulent fields in incompressible plasmas.
The set of equations which is solved in the MHD code GISMO are the incompressible MHD equations
\begin{align}
 \pa{\vec{u}}{t} &= \vec{b} \cdot \nabla \vec{b} -\vec{u} \cdot \nabla \vec{u} -\nabla P + \nu \nabla^{2h} \vec{u}   \\
 \pa{\vec{b}}{t} &= \vec{b} \cdot \nabla \vec{u} -\vec{u} \cdot \nabla \vec{b} + \nu \nabla^{2h} \vec{b}\label{eq:mhdset}
\end{align}
with the magnetic field $\vec{b}\equiv \vec{B}/\sqrt{4\pi \rho}$ with constant mass density $\rho$ and the fluid velocity $\vec{u}$. The total pressure is denoted by $P$ and describes both, thermal and magnetic pressure with $P=p+B^2/(8\pi)$. Viscous and Ohmic dissipation are given by the generalised resistivity $\nu$, which causes wavenumber-diffusion. We consider here also hyperdiffusivity, which occurs for $h>1$ Especially for the fast solar wind, which we are interested in, this fluid can be considered as incompressible. This leads together with the solenoidality condition for the magnetic field to the boundary conditions
\begin{align}
 \nabla \cdot \vec{u} &= 0  \\
 \nabla \cdot \vec{b} &= 0
\end{align}
Using these boundary conditions, it is possible to find a closure for the MHD equations. The pressure $P$ may be derived by taking the divergence of the MHD equations. This in turn yields \cite{marongold}
\begin{align}
 \nabla^2 P = \nabla \vec{b} : \nabla \vec{b} -\nabla \vec{u} : \nabla \vec{u}. \label{eq:pressureclosure}
\end{align}

The solution for incompressible fluid problems can be achieved by the spectral method.

In the incompressible regime of a magnetised plasma the MHD-turbulence consists of only two types of waves, which propagate along the parallel direction - the so-called pseudo- and shear Alfv\'en waves. First ones are the incompressible limit of slow magnetosonic waves and play a minor role within anisotropic turbulence \cite{marongold}. The pseudo Alfv\'en waves polarisation vector is in the plane spanned by the wavevector $\vec{k}$ and $\vec{B_0}$. The shear waves are transversal modes with  a polarisation vector perpendicular to the $\vec{k}$ - $\vec{B_0}$ plane. They are circularly polarised for parallel propagating waves.
Both species exhibit the dispersion relation $\omega^2=(v_A k_\parallel)^2$.

Since the model consists only of these two wave types it is suitable to use a description with Alfv\'enic waves moving either forwards or backwards. This is achieved by introducing the Els\"asser variables \cite{elsasser}
\begin{align}
 \vec w^- &= \vec v + \vec b - v_A \vec e_\parallel \nonumber \\
 \vec w^+ &=\vec v - \vec b + v_A \vec e_\parallel,
\end{align}
and transforming the Eqs. \ref{eq:mhdset} into a suitable form of
\begin{align}
  \left(\partial_t - v_A k_z\right) \tilde w_\alpha^\mp &= \frac{i}{2} \frac{k_\alpha k_\beta k_\gamma}{k^2} \left( \widetilde{w_\beta^+ w_\gamma^-} +  \widetilde{w_\beta^- w_\gamma^+}\right) \nonumber \\
		                               &-ik_\beta \widetilde{w_\alpha^\mp w_\beta^\pm} - \frac{\nu}{2} k^{2h} \tilde w_\alpha^\mp \nonumber \\
  k_\alpha \tilde w^\pm_\alpha = 0
\end{align}

\begin{figure}[ht]
  \includegraphics[height=7cm]{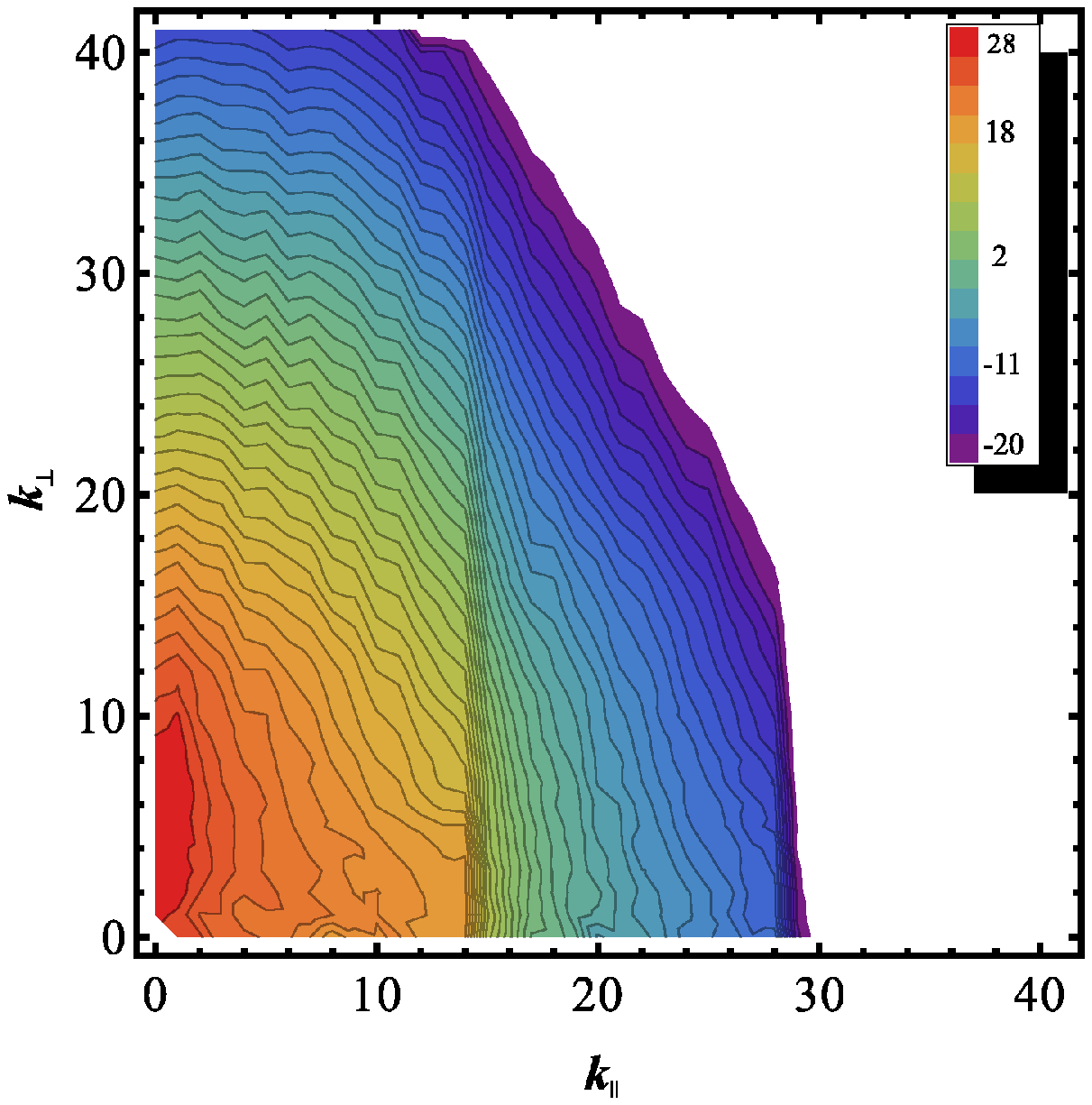}\hfill\includegraphics[height=7cm]{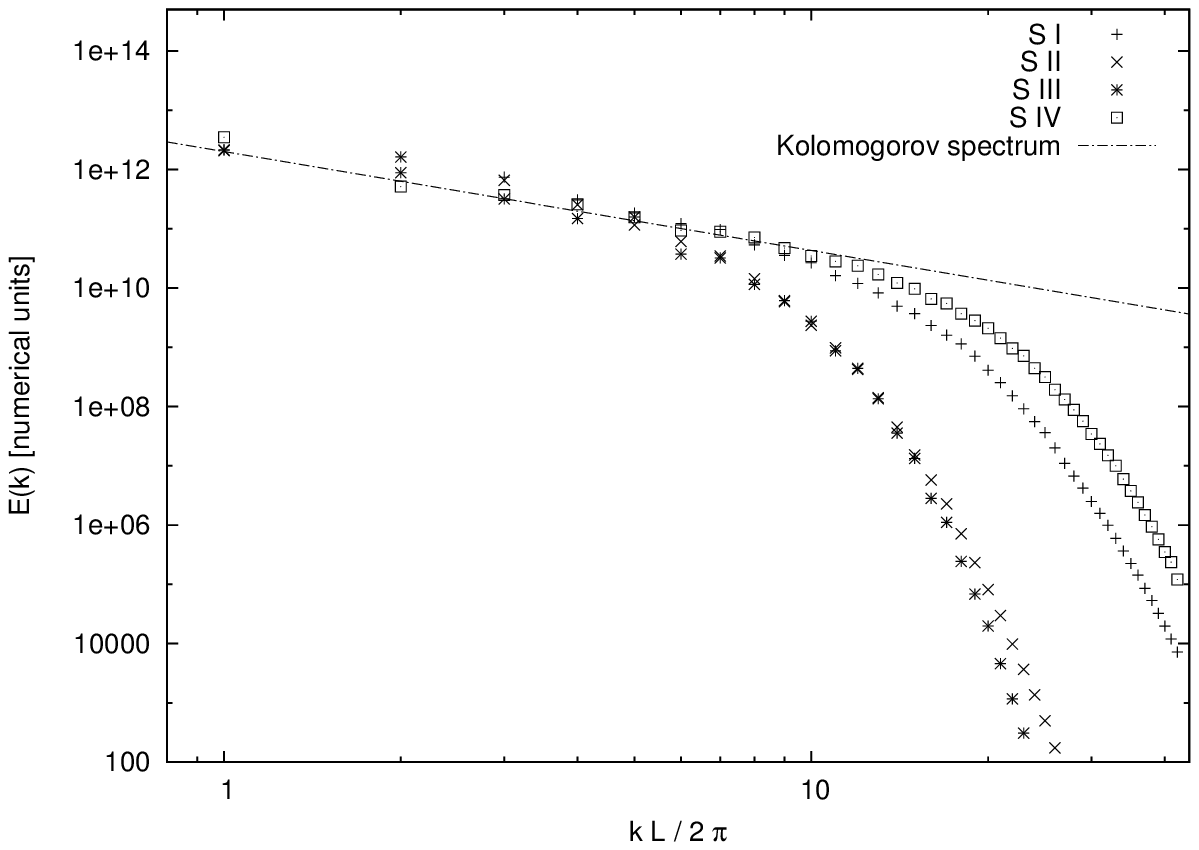}
  \caption{Results of the background simulation performed with GISMO. The left figure shows a two dimensional energy spectrum, that clearly shows signs of Goldreich-Sridhar cascade \citep{gsrev} in which energy is transport preferentially along $k_\perp$ until a critical balance is reached, after which energy can also be transported along $k_\parallel$. The figure on the right side shows the corresponding one-dimensional spectrum, which is approximately a $k^{-5/3}$ spectrum.}
  \label{fig:gismoback}
\end{figure}
\begin{figure}[ht]
  \includegraphics[height=7cm]{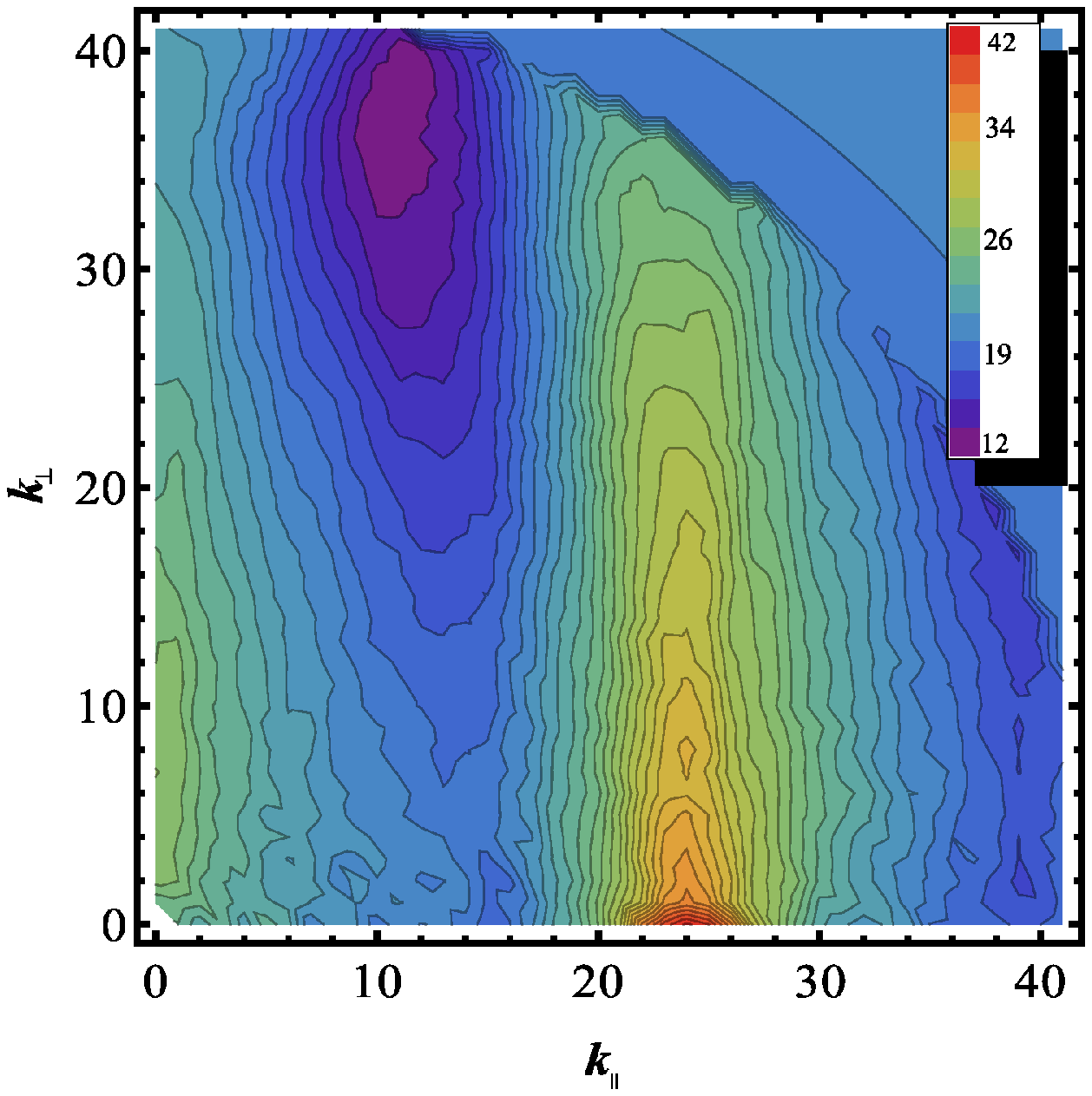}\hfill\includegraphics[height=7cm]{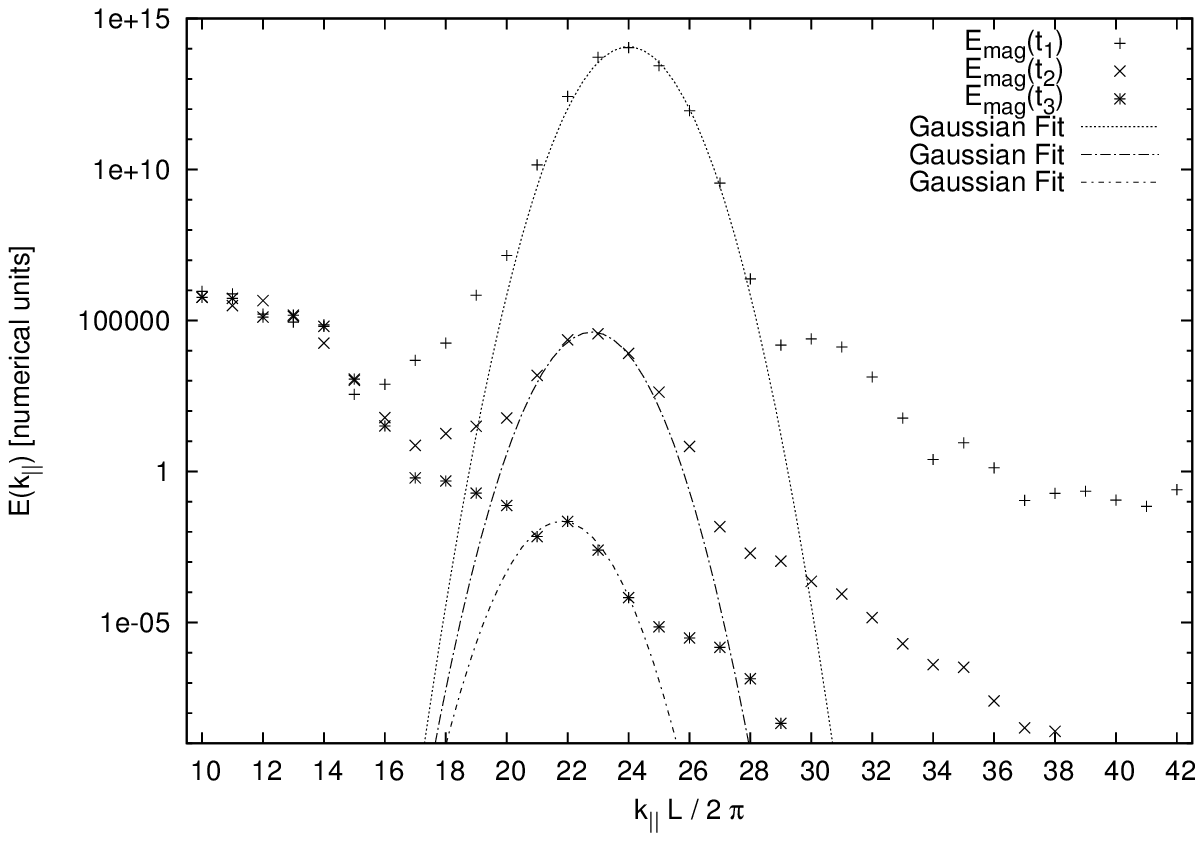}
  \caption{Results of a simulation performed with GISMO in which energy is injected at $k=24$ starting from a background simulation. The left figure shows a two dimensional energy spectrum, where the transport in $k_\perp$ direction also for the injected energy can be seen. The figure on the right side shows the corresponding spectrum in parallel direction, where the added bump from he injected energy can be seen. This is approximately in the dissipation regime of the Kolmogorov spectrum.}
  \label{fig:gismobeam}
\end{figure}

Two important scenarios with regard to the turbulence simulation have been considered:
\begin{itemize}
  \item A scenario in which energy is continuously injected at smallest wavenumbers until an equilibrium of driving and dissipation leads to a turbulent inertial range. This is the background simulation (cf. fig. \ref{fig:gismoback} on which all other simulations build up. As a variety the decaying turbulence, in which the driving is turned off is also studied
  \item Another important scenario is the injection of energy at medium wave numbers, which resembles the energy injected into the plasma by energetic protons (cf. fig. \ref{fig:gismobeam}). This scenario is interesting in terms of physics since proton beams are one important source of energy, but they are also numerically interesting since the wave number space local turbulence ratio can reach values of $\delta B/B_0 > 1$
\end{itemize}

Into the fields, generated by the MHD code, charged test-particles are injected. The Lorentz force
\begin{align}
 \frac{\text{d}}{\text{d}t} \; \gamma \, \vec{v} = \frac{q}{m c}\left[ \,c \vec{E}(\vec{x},t) \,+ \, \vec{v} \times \vec{B}(\vec{x},t)\, \right], \label{eq:lorentzforce}
 \end{align}
is acting upon the particles. Here the electric field is generated from the MHD fields assuming an ideal Ohm's law with $\vec{E} = -\vec{u}\times\vec{B}=-1/4(\vec{w}^- + \vec{w}^+)\times(\vec{w}^+ - \vec{w}^-)$

A suitable numerical approach for solving Eq. (\ref{eq:lorentzforce}) for gyrating particles is the implicit scheme of the \emph{Boris-push}.
The basic idea has been given by \cite{boris70} where the iterations of the Lorentz force are separated in two partial steps.
First, the particles are accelerated by the electric field within a half time step. Second, the gyromotion of the particles is calculated, which is caused by the magnetic field.
After that, the electric fields acts again for another half time step to complete the iteration. This approach leads to a discretisation of the Lorentz force (for the detailed set of Eqs. see \cite{langdon}).

The advantage of the Boris-push is the very high numerical stability. The particles are assumed to undergo gyromotions, hence the particle orbits themselves are stable for an arbitrary time discretisation. Even in the limit of $\Delta t^\text{num} \gg \Omega^{-1}$ the particle orbit is stable, but converges to an adiabatic drift motion. The limitation of this method is the correct resolution of the Larmor radius $r_L$. If the timestep is chosen too large, this would lead to a big deviation from the analytical $r_L$. \textsc{Gismo--Particles} measures the deviation from $r_L$ and adapt it to the preferred value. To specify, in our simulations an accuracy of the order of $|r_L-r_\text{measured}|/r_L \approx 10^{-5}$ was used.

A limitation to the method of the Boris-push are ultrarelativistic particle speeds. In this case the conservation of energy would be violated, since the ideal
ohmic law is not fulfilled anymore. Beyond Lorentz factors of $\gamma \approx 10^3$ fictitious forces start to act and this approach is not applicable
furthermore \cite{vay08}.

Both parts of \textsc{Gismo} are calculated for each step. After iterating the Els\"asser MHD-fields $\vec w^\pm$, they are transformed into the physical electric and magnetic fields which are transferred to \textsc{Gismo--Particles}. Then the Boris-push will be performed. Each particle will respond to its local fields, which are calculated by an averaging method via three-dimensional splines \cite{2011ASTRA...7...21S,2012ApJ...750..150W}. Periodic spatial boundary conditions were used, thus the number of particles remained constant in each simulation.



\end{document}